# Strong Quasiparticle Trapping In A 6x6 Array Of Vanadium-Aluminum Superconducting Tunnel Junctions

G. Brammertz[a], A. Peacock[a], P. Verhoeve[a], A. Kozorezov[b],
R. den Hartog[a], N. Rando[a], R. Venn[c]

[a] *Space Science Department, ESA/ESTEC, P. O. Box 299, 2200AG Noordwijk, The Netherlands.*
[b] *Department of Physics, Lancaster University, Lancaster LA1 4YB, UK.*
[c] *Cambridge MicroFab Ltd., Trollheim Cranes Lane, Kingston, Cambridge CB3 7NJ, UK.*

**Abstract.** A 6x6 array of symmetrical V/Al/AlOx/Al/V Superconducting Tunnel Junctions (STJs) was fabricated. The base electrode is a high quality epitaxial film with a residual resistance ratio (RRR) of ~30. The top film is polycrystalline with an RRR of ~10. The leakage currents of the 25x25 $\mu m^2$ junctions are of the order of 0.5 pA/$\mu m^2$ at a bias voltage of 100 $\mu V$, which corresponds to a dynamical resistance of ~ 3 $10^5$ $\Omega$. When the array was illuminated by 6 keV X-ray photons from a $^{55}$Fe radioactive source the single photon charge output was found to be low and strongly dependent on the temperature of the devices. This temperature dependence at X-ray energies can be explained by the existence of a very large number of quasiparticle (QP) traps in the Vanadium. QPs are confined in these traps, having a lower energy gap than the surrounding material, and are therefore not available for tunneling. The number of traps can be derived from the energy dependence of the responsivity of the devices (charge output per electron volt of photon input energy).

As a process-route precursor to the fabrication of low $T_c$ STJs based on Mo and Al for use as optical to X-ray detectors in Astronomy, V-Al-AlOx-Al-V STJs were produced. Their characteristics from a film quality, leakage and spectral performance point of view are presented in this paper.

The multilayer is deposited on a sapphire substrate without breaking vacuum. First a 100 nm thick epitaxial V film is sputtered at 550° C in an UHV chamber. After cooling the sample a 30 nm thick Al film is deposited on top of the base V. 10 Å of Al are then oxidized to form the AlOx barrier of the junction. On top of this barrier another 30 nm of Al and 100 nm of V is deposited. The trilayer wafer then is removed from the UHV chamber and a ~7 nm thick VOx film naturally forms on top of the multilayer. Figure 1a shows a TEM image of the multilayer.

Despite considerable lattice mismatch between the substrate and the V film, TEM has shown the base electrode to comprise an epitaxial film with a residual resistance ratio $\rho_{293K}/\rho_{10K}$ (RRR) of ~30, as opposed to the top film which is polycrystalline with an RRR of ~10. This corresponds to a mean free path $\lambda_{10K}$ in the base V of ~120 nm and a normal state diffusion constant $D_{10K}$ of ~60 $cm^2$/sec. The polycrystalline top film has a 3 times lower mean free path and diffusion constant. The junction's geometry is defined by using a photolithographical etching mask. Two different mask layouts are available. One is a 6x6 array with junctions having a 25 µm side length. The other layout is a science chip with junctions of different side lengths ranging from 7 to 70 µm. After the different etching steps the junctions are covered with an insulating layer. Then Nb top contacts and base contact plugs are deposited through vias in the insulating layer. Figure 1b shows a Nomarski microscope image of the array layout. One can see that the base material was not completely etched through everywhere and that 8 junctions in the lower half of the array are still interconnected via this residual base material.

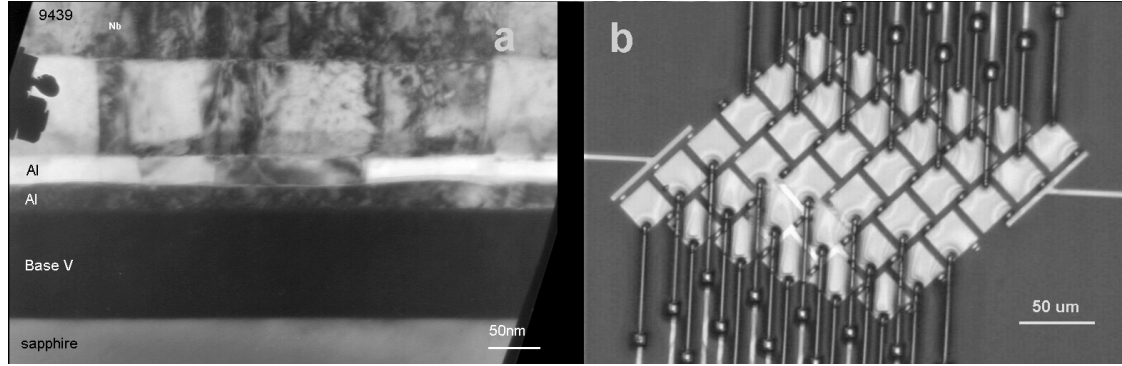

**FIGURE 1. a**: TEM image of the multilayer. The base V is clearly epitaxial, whereas it is not clear if the base Al is epitaxial or polycrystalline, as there are strong contrasts in the film. The top Al and V films are polycrystalline with columnar grain sizes of ~ 100 nm. On top of the trilayer a Nb film can be seen, which is deposited for the contacts. **b**: Nomarski microscope image of the 6x6 V/Al array. The junction side length is 25 µm. In the lower half of the array 8 junctions are interconnected through residual unetched base material.

The devices were operated at 300 mK in a $^3$He cryostat, where IV curves and spectra can be acquired. The junctions presented here are low leakage junctions, although a small residual leakage is still present. Figure 2a shows an IV-curve of a typical device of the array. The leakage current of such a junction at a bias of 100 µV is 0.5 pA/µm$^2$, which corresponds to a dynamical resistance $R_d \sim 3 \cdot 10^5$ Ω. The normal resistance of the same junction is $R_n \sim 0.2$ Ω. This gives a quality factor $R_d/R_n \sim 1.5 \cdot 10^6$. The energy gap of the 100 nm V- 30 nm Al bilayer is $\Delta_g = 530$ µeV, compared to that of bulk V: $\Delta_{g,b} = 820$ µeV.

All junctions were illuminated with an $^{55}$Fe source ($E_{K\alpha}$=5895 eV; $E_{K\beta}$=6490 eV). If we assume that the energy gap is reduced throughout the device and that the number of initial QPs created by the absorption of a photon can be determined using the same equation as in Nb and Sn [1,2]: $Q_0 = E/1.75 \Delta_g$, we obtain an initial number of QPs $Q_0 \sim 6.4 \cdot 10^6$. Our photon detection experiments revealed however a charge output of the order of 3-4 $10^6$ e$^-$. This means that the mean number of tunnels <n> in our V STJs is as low as 0.5. Figure 2e shows for example the uniformity of the charge output of the array at a bias voltage of 250 µV. One clearly sees that all devices show this low charge output. The devices with zero charge output in the front of the figure are the junctions connected through the residual base metal. For these junctions most QPs created in the base film diffuse out before they can tunnel and as a consequence the charge output is even lower. Another interesting feature observed is that the charge output ratio between the peaks of the K$_\beta$ and K$_\alpha$ lines $Q_{K\beta}/Q_{K\alpha}$ is equal to 1.12, instead of the energy ratio of the two lines $E_{K\beta}/E_{K\alpha}$, which is 1.10. In Nb and Ta based STJs this charge output ratio is usually lower or equal to 1.10 [3-5] because of non-linear QP self-recombination effects, which enhance losses at higher QP densities [6]. Our V/Al junctions are thus not limited by QP self-recombination despite the high densities we are dealing with. In order to explain the very low charge output and the unexpected non-linearity behavior we simulated a photon detection experiment in a 10 μm side length junction with a bias voltage of 260 μV applied using a model developed by Poelaert et al [7,8]. The model includes effects from QP self-recombination, QP trapping in local traps and QP de-trapping due to phonon absorption or QP recombination. The numerical values for all parameters used for the simulation can be found in table 1. We choose a 10 μm junction for our simulations because of the better energy resolution in the smaller devices (figure 2c). The 170 eV FWHM energy resolution in the 10 μm device allows us to discern the $K_\alpha$ and $K_\beta$ lines, whereas the 25 μm devices with an energy resolution of 500 eV do not separate these lines. The charge outputs of all devices are similar ($<n> < 1$).

**TABLE 1. Parameters for the photon absorption simulation**

| Parameter (Symbol) | Value (Unit) | Parameter (Symbol) | Value (Unit) |
|---|---|---|---|
| QP balance energy ($E_0$) | 707 (μeV) | Cooper pair breaking time ($\tau_{PB}$) | 69 (psec) |
| Recombination coefficient (R) | 307 (μm$^3$/sec) | Tunneling time ($\tau_{tun}$) | 170 (nsec) |
| Diffusion constant base ($D_B$) | 30 (cm$^2$/sec) | Diffusion constant top ($D_t$) | 10 (cm$^2$/sec) |
| Phonon transmission through barrier ($T_b$) | 0.3 | Phonon transmission to substrate ($T_s$) | 0.2 |
| Phonon loss time ($\tau_{PL}$) | 1 (nsec) | De-trapping efficiency (η) | 0.3 |
| QP loss time in base ($\tau_{l,b}$) | 0.84 (μsec) | QP loss time in top ($\tau_{l,t}$) | 0.88 (μsec) |
| QP trap density base ($N_{t,b}$) | 77000 st./μm$^3$ | QP trap density top ($N_{t,t}$) | 82000 st./μm$^3$ |
| Trapping coefficient ($c_t$) | 8000 μm$^3$/sec | | |

The first 4 parameters from the table were derived by using Golubov's et al. equations of characteristic times [9] applied to a proximized density of states calculated in [10]. The diffusion constants in the superconducting state were derived from the normal state diffusion constants. The next 4 parameters are typical values taken from [8]. The last 5 parameters from the table were varied in order to fit the numerical results to the 8 experimental points (Charge output and rise time of the pulse in the top and base film for the $K_\alpha$ and the $K_\beta$ lines) (figure 2d). From these simulations it is clear that the large number of QP traps is responsible for the low charge output in our detectors. The number of QP traps $N_t$ is one to two orders of magnitude higher than in Nb and Ta films. Also the trapping efficiency $c_t$ is one to two orders of magnitudes higher, which makes the product $c_t N_t$ respectively 3 and 4 orders of magnitude higher than in Nb and Ta STJs [8]. In future work spectra with different photon energies have to be acquired in order to extend the data range.

Another experimental result that supports this strong trapping theory is the strong dependence of charge output on temperature (figure 2b). The increase in charge output at higher temperatures is explained by the enhanced de-trapping mechanism due to phonon absorption, which is of course more effective at higher temperatures. Strong temperature dependencies of the charge output have already been observed in Ta based STJs [11], but these results were obtained at optical wavelengths. At higher photon energies the temperature dependence for such Ta based devices vanishes because the number of traps is no longer of importance when compared to the number of QPs in the junction. In these V STJs however the trap density is so high that the temperature effect can still be seen for photon energies as high as 6 keV. A maximum charge output is obtained at 750 mK. Above this photon energy recombination with thermal QPs become more important and the charge output falls.

In conclusion, the characteristics of low leakage V/Al STJs used as photon detectors were presented. For X-ray photons the junctions appear to have very low charge output ($<n> < 1$), which depends strongly on photon energy and temperature. Numerical simulations of photon detection in such systems show that the reason for this behavior is the large number of QP traps (regions of slightly suppressed energy gap) and the high trapping efficiency in the V.

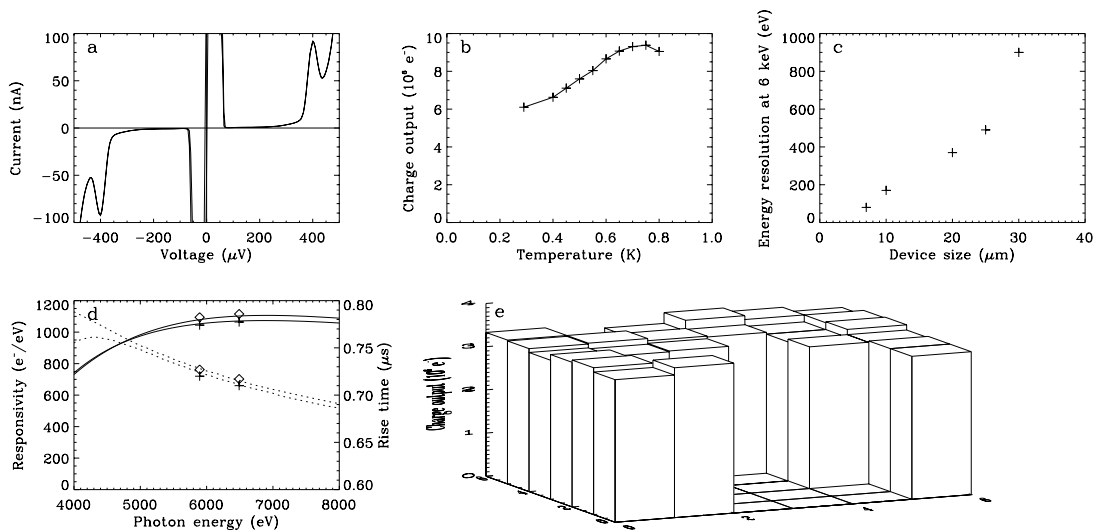

**FIGURE 2. a**: IV-curve of a typical junction from the 6x6 array. **b**: Charge output dependence on detector temperature of the 10 µm junction with a bias voltage of 250 µV applied. **c**: Energy resolution of the base film of the V/Al STJs as a function of device size. **d**: Simulated responsivity (solid lines) and rise-time (dashed lines) of the integrated pulses from the top and base films as a function of photon energy. The experimental points for the $K_\alpha$ and $K_\beta$ lines of the 10 µm junction are also shown (top film: diamonds, base film: crosses). **e**: Charge output in $10^6$ e$^-$ for all devices of the 6x6 array. The pixels with no charge output in the front are the pixels interconnected through residual base metal.